# Dynamic Nuclear Polarization in Battery Materials

Shira Haber [(1)] and Michal Leskes [(1)]*

(1) Department of Molecular Chemistry and Materials Science, Weizmann Institute of Science, Rehovot Israel

*michal.leskes@weizmann.ac.il

**Abstract**

The increasing need for portable and large-scale energy storage systems requires development of new, long lasting and highly efficient battery systems. Solid state NMR spectroscopy has emerged as an excellent method for characterizing battery materials. Yet, it is limited when it comes to probing thin interfacial layers which play a central role in the performance and lifetime of battery cells. Here we review how Dynamic Nuclear Polarization (DNP) can lift the sensitivity limitation and enable detection of the electrode-electrolyte interface, as well as the bulk of some electrode and electrolyte systems. We describe the current challenges from the point of view of materials development; considering how the unique electronic, magnetic and chemical properties differentiate battery materials from other applications of DNP in materials science. We review the current applications of exogenous and endogenous DNP from radicals, conduction electrons and paramagnetic metal ions. Finally, we provide our perspective on the opportunities and directions where battery materials can benefit from current DNP methodologies as well as project on future developments that will enable NMR investigation of battery materials with sensitivity and selectivity under ambient conditions.

**TOC**

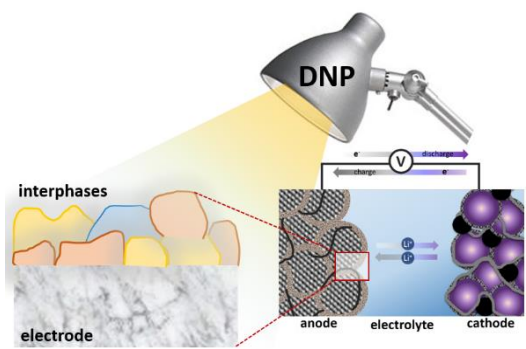

# 1. Introduction

Rechargeable batteries and in particular the lithium-ion battery (LIB), have revolutionized our life by enabling the development of portable electronic devices.[1,2] In recent years, it became clear that batteries are essential to make the transition towards utilization of sustainable and clean energy resources: from powering electric vehicles to storing energy, originating from transient sources such as solar and wind, so that it can be used efficiently on a smart grid.[3,4] The implementation of rechargeable batteries in these large-scale applications requires significant improvements in current batteries' systems. Advancements are needed on all fronts, depending on the application, and include higher energy or power capabilities of the battery packs, increased safety and longer lasting storage systems. Thus, to accommodate these needs, there have been enormous efforts of research and development in both academia and industry on all levels, from materials design to battery management.[5,6]

Analytical characterization tools are an essential ingredient in the development of new and improved battery materials and cells. Analysis of materials prior, during, and post operation, and understanding of their failure mechanisms are required for rationally designing energy storage materials. Among the common tools used by materials chemists and engineers, such as X-ray diffraction, electron microscopy and X-ray photoelectron and optical spectroscopies, solid state NMR (ssNMR) spectroscopy has emerged as a powerful approach. To the NMR spectroscopist the advantages are clear: NMR provides detailed, quantitative and chemically specific insight into the composition, structure and dynamic properties of the materials. Such atomic-molecular scale information is extremely beneficial for understanding the functioning and failure mechanisms in battery materials and is used to guide the design of improved materials systems.[7,8] Nevertheless, the inherent low sensitivity of ssNMR often impedes its application in the study of interfaces and interphases which, as will be discussed below, play a central role in the performance of rechargeable batteries.

In this trends article we aim to describe how Dynamic Nuclear Polarization (DNP) can be integrated in the study of battery materials. While magic angle spinning (MAS) DNP has been increasingly employed in materials science,[9] its application to battery materials is relatively new and seems like a natural application for the approach. Nevertheless, battery materials are often associated with properties such as conductivity, paramagnetism, chemical reactivity and are often functional as a composite of organic and inorganic components. These properties make them distinct from many of the applications of MAS-DNP to date. As MAS-DNP commonly involves addition of a solution of organic radicals to the material and measurements at cryogenic temperatures at limited MAS frequencies, its application to battery materials is not a trivial extension of the current approaches, as will be described in this article. Clearly the implementation of DNP can help gain sensitivity and alleviate some of the limitation of ssNMR. However, beyond sensitivity, recent developments in DNP methodology offer new opportunities to probe materials and their interfaces with selectivity in addition to sensitivity. These can have significant impact on our understanding of battery materials, which will enable the development of better storage systems.

We will first describe the basic components of a typical rechargeable battery cell followed by an overview of the current materials challenges and developments which are relevant for studies by NMR spectroscopy and where DNP can have an impact. We will then detail the magnetic resonance properties of battery materials in the context of their study with ssNMR and DNP. Next, we will provide an overview of the DNP approaches using various polarizing agents, the relevant DNP mechanisms and the benefits and limitations of the different approaches. We will describe how these approaches were implemented in battery research and the type of information that was gained. Finally, we will describe the future challenges and opportunities for DNP on battery materials.

## 2. Rechargeable Batteries: Overview and Current Challenges

A battery cell is made of two electrodes, a cathode and an anode, differing in their redox potential and separated by an ion conducting electrolyte,[2] as schematically shown in Figure 1a. In most commercial lithium-ion cells, the anode is made of graphite and the cathode of a transition metal oxide (typically a layered oxide such as $LiTMO_2$ with TM being a mixture of $Co^{3+}$, $Mn^{3+}$ and $Ni^{3+}$). The common liquid electrolyte consists of a lithium salt dissolved in organic solvents and is made of 1M $LiPF_6$ solution in a 1:1 ratio of ethylene carbonate (EC) and dimethyl carbonate (DMC) (and possible low percent of organic additives). In practice, the electrodes are made of a composite matrix containing mostly the active material (the redox active component), conductive carbon additive, to increase the electrical conductivity, and a polymer binder. The composite is deposited on a metal current collector (commonly copper for anodes and aluminium for cathodes) to create a uniform electrode film. The two electrodes are physically separated by a membrane (made of polypropylene or borosilicate), which is soaked by the electrolyte. For the battery cell to function the electrolyte has to reach the active particles by wetting the electrode and enabling efficient ion transport between the two electrodes through the separating membrane.

The main benchmarks used for evaluating battery cells are: (i) *Energy density,* which is determined by the redox potential difference between the active materials in the electrodes and the capacity of the electrodes, (ii) *power density*, namely how fast a cell can be discharged/charged while maintaining sufficient capacity, (iii) *cycle life*, how many cycles a LIB can sustain without significant loss in performance and (iv) *safety*. These are all inherent properties of the materials used in the battery cell and the chemical and electrochemical interactions between them.

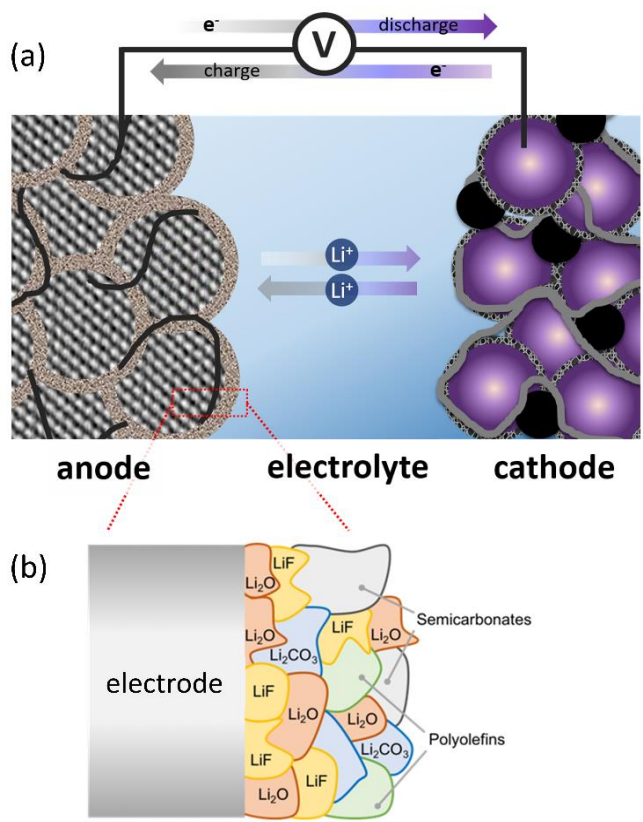

Understanding how the materials' structure and properties affect their performance in the battery cell is often far from trivial. The active materials themselves can undergo structural changes during the redox process. These can be due to the mechanism by which the materials store charge[5] or due to surface transformations such as oxygen evolution, transition metal dissolution into the electrolyte and loss of active

**Figure 1** (a) Schematic representation of a battery cell made of composite anode and cathode separated by a liquid electrolyte. The composites are typically made of a polymeric binder (black and grey strands), carbon black (black circles) and active material (grey and purple particles). (b) Reactions with the electrolyte result in deposition of solid phases at the surface of the electrodes, the SEI. The figure depicts a typical composition of the SEI forming on LIB anodes.

material.[10] Furthermore, chemical and electrochemical reactions between the components are almost unavoidable and their extent depends on the materials used. As most anode materials, graphite, Si or lithium metal, operate at low potentials compared to the stability of the electrolyte components, they are susceptible to electrolyte degradation at the surface of the electrodes. This leads to accumulation of interphases, organic and inorganic phases deposited at the electrode-electrolyte interface. Thus, the performance of LIB cells depends on the formation of a stable passivation layer on the electrode-electrolyte interface called the solid-electrolyte interphase (SEI, Figure 1b).[11–13] This layer, which ideally should form at the initial cell discharge-charge cycles, should maintain efficient lithium ion transport and prevent other non-reversible reactions from taking place on the electrode surface. The interfacial properties affect the capacity, power, and lifetime of the LIB, thus by passivating the surface and providing efficient ionic conduction pathways, beneficial SEI leads to enhanced electrochemical performance.[13] It is clear, that determining the chemical composition and
44

structure of these interphases is essential for understanding the functionality of the SEI on the anode and the cathode. Insofar, diagnosis is key to the development of new and improved materials and interfaces.

Current research efforts are invested in all components of the cells (electrodes and electrolyte). These range from design of new materials or new chemistries for energy storage coupled with efforts to understand their electrochemical function and interactions. While we do not intend to cover all of the new developments in the field, we will provide a short overview of the main developments which will be relevant for the discussion later.

For example, to increase the energy density of the cell, a common approach is to introduce high energy cathodes that operate above 4.2 V (vs. Li) and/or provide capacity exceeding 140 mAh/g, which are commonly obtained with $LiCoO_2$. Such increase in energy also leads to a plethora of surface reactions on the cathode side. Due to lack of stable SEI on the cathode, there is need for developing surface treatments that will passivate the cathode surface, for example through atomic or molecular layer deposition.[14] On the anode side, researchers are looking for alternatives which will surpass the 372 mAh/g obtained with graphite. For example, Li alloying materials such as Si can result in a 10-fold increase in capacity. However, the large volume increase in lithiated Si leads to continuous SEI formation. This can be controlled to some extent through design of artificial SEI, which can accommodate the volume expansion, or the use of nanosize Si particles.[15,16]

Li metal is also considered as a high energy alternative, yet it leads to uncontrollable SEI formation, non-uniform Li deposition and formation of lithium dendrites. Such dendritic structures are a safety hazard when they cause a short circuit in the cell in the presence of flammable organic electrolyte. To overcome safety issues associated with liquid electrolytes, solid state electrolytes (SSE) are being developed, consisting of ceramic or polymeric solid ion conductors or a composite of both.[17,18] SSE raise many scientific and engineering challenges associated with limited conductivity in the solid state at room temperature, interfacial stability and wettability of the electrodes.

Other directions to improve battery cells involve development of new charge storage chemistries. Layered electrode structures (such as $LiCoO_2$ and graphite) store lithium reversibly through intercalation. Development of conversion chemistries, where a metal oxide/sulfide/fluoride convert to the corresponding lithium oxide/sulfide/fluoride and metallic particles, have potentially high capacity yet their reversibility is limited.[19] Another route for increasing the capacity involves the use of metallic lithium anodes with sulfur or oxygen cathodes. These involve complex reactions in which, sulfur and oxygen are reduced to form lithium sulfides and oxides, respectively, within a porous carbonaceous cathode. While in the Li-S system the main challenge to overcome is associated with shuttling of polysulfide chains through the electrolyte to the lithium metal,[20] Li-O$_2$ cells suffer from significant electrolyte and electrode decomposition processes.[21]

Finally, the expected increase in demand for rechargeable batteries raises the need for cheap and abundant alternatives to lithium. Sodium ion batteries (SIB) are developing as a suitable alternative to LIB, in particular for grid storage where the low cost and abundance of sodium are an advantage while its relative lower energy density is less of a concern.[22] Other options are development of cells based on multivalent ions such as $Al^{3+}$, $Mg^{2+}$ and $Ca^{2+}$. These new chemistries call for significant research efforts in identifying suitable electrodes and electrolyte materials along with insight into their electrochemical performance and interfacial chemistry.

## 3. Magnetic Resonance Properties and Characterization of Battery Materials

In the past decades ssNMR spectroscopy has been extensively applied and utilized to study LIB and other emerging battery materials.[7,23–30] There are several excellent recent reviews on the topic and here we aim to highlight the key capabilities of ssNMR spectroscopy, as well as identify the main questions where DNP can contribute to the investigation of battery materials. We place emphasis on the different electronic, magnetic and chemical properties of the materials, and how these affect their study by NMR. These properties will determine whether such materials are amenable to DNP in its current formulations, and allow us to highlight some of the directions that can be explored to expand the applicability of DNP to such materials.

NMR is particularly insightful in elucidating the electrochemical mechanism by which electrode materials store energy. It can be applied to study whole battery cells either in-situ (stopping the electrochemical process) or operando (simultaneously measuring NMR spectra with electrochemical cycling).[7,31] Such measurements are limited by resolution as they are most commonly employed to static cells, although recently it was demonstrated that in-situ measurements can be done with MAS by assembling the battery cell in the NMR rotor.[24] High resolution MAS spectra are collected through post-mortem ex situ measurements which require battery disassembly in argon environment and removal of the electrode material from the aluminum\copper current collector before testing.

The sensitivity, resolution and ability to employ complex NMR pulse sequences, and hence the level of detail that can be gained from the different measurements, strongly depend on the nature of the studied materials. These vary significantly across the different battery components as elaborated below.

*Paramagnetic electrodes:* Most cathode materials are transition metal oxides, fluorides or phosphates which display varying degree of paramagnetism. Thus, NMR spectra of electrodes acquired through detection of $^{6,7}$Li in LIB or $^{23}$Na in SIB, are dominated by the interactions between the unpaired *d* electrons in the TM and the nuclei. Fermi contact shifts lead to unique resonance frequencies that can be used to follow the electrochemical process through the shift sensitivity to the TM oxidation state. On the other hand, anisotropic dipolar interactions lead to severe spectral broadening which limit the ability to identify isotropic resonances. These strong paramagnetic interactions also result in very short longitudinal relaxation ($T_1$) times in the µs-ms range. Resolution can be gained by using low magnetic fields, fast MAS frequencies and sideband separation sequences such as pj-MATPASS and aMAT (Figure 2).[32,33] With these approaches, ssNMR spectra provide detailed insight into the electrochemical and structural changes in the bulk of paramagnetic electrode materials.

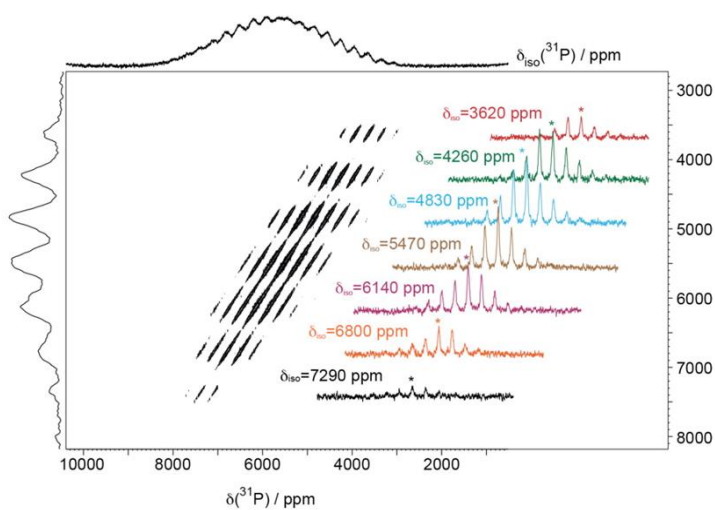

**Figure 2** Two-dimensional $^{31}$P aMAT experiment performed on LiFe$_{0.5}$Mn$_{0.5}$PO$_4$ in order to separate the isotropic $^{31}$P resonances and corresponding sideband manifolds. The top projection shows the overlapping sideband patterns in the one-dimensional $^{31}$P spectrum while the projection in the indirect dimension contains only the isotropic shifts. The seven resolved sideband patterns are also shown. Measurements were performed on 11.75 T (500 MHz) spectrometer at 60 kHz MAS. Adapted from ref. 32 with permission.

*Conductive electrodes and additives*: Many anode materials and some cathodes are metallic in nature or can become metallic upon (de)lithiation. Some examples include lithiated graphite, elemental metal anodes as well as delithiated LiCoO$_2$ and the family of Li$_2$RuO$_3$ cathodes. In these electrodes, interactions with the conduction electrons dominate the NMR spectra, leading to Knight shift of the nuclear resonances and fast relaxation times. However, in highly metallic materials the detected resonances are predominantly isotropic which results in high resolution spectra even under static in situ/operando conditions. As such, ssNMR measurements provide useful insight into the evolution of metallic species. In particular NMR is a powerful approach for detecting and quantifying metallic dendrites as their dimension is smaller than the skin depth of the RF irradiation.[34–36]

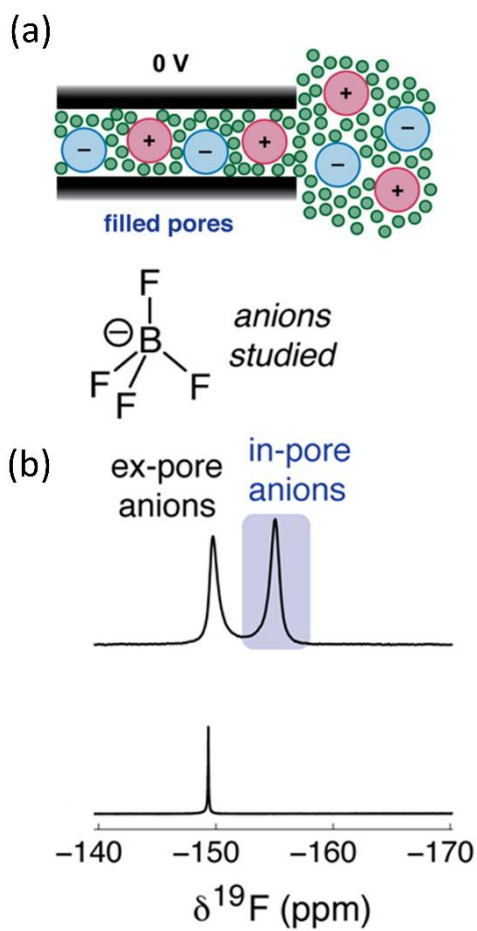

**Figure 3** (a) Schematic illustration of filled carbon pores. The carbon (slit-) pore walls are represented by black rectangles. (b) $^{19}$F NMR (9.4 T) measurements of YP50F activated carbon soaked with a typical supercapacitor electrolyte (top), NEt$_4$–BF$_4$/d-acetonitrile (1.5 M), and the neat electrolyte (bottom) recorded with MAS at a frequency of 5 kHz. Adapted from ref 41 with permission.

Many electrode materials are made of carbonaceous materials, either as the active electrode material (graphite for LIB, hard carbon for SIB), as a scaffold material (mesoporous carbons, carbon nanotubes, carbon black etc. in Li-S and Li-O$_2$ cathodes) or as conductive additives (carbon black and nanotubes/wires). These are typically semi-conductive and do not lead to Knight shifts but pose a challenge in NMR measurements due to RF absorption which lead to sample heating, attenuation of the effective RF power as well as significant shortening of nuclear relaxation times. Furthermore, they may induce strong ring currents which can be used to identify adsorbed species through shifts in their NMR resonance frequency (Figure 3).[37–41]

*Composite materials and Interphases*: Most electrodes are composites made of the active inorganic powder, an organic polymer binder (typically Teflon or polyvinyl fluoride, PVDF) and conductive carbon additive. Composite solid electrolytes are made of a mixture of inorganic ceramic particles, polymer species (commonly polyethylene oxide) and a lithium salt. As mentioned above, the SEI layer and other artificially formed interphases (coatings) are complex structures made of mixtures of organic and inorganic phases. Thus, battery materials are often very heterogeneous and disordered. Interphases are commonly associated with diamagnetic properties, and their NMR response is dominated by chemical shifts (typically across a much narrower frequency range than in paramagnetic species) and broadening due to heterogeneity. Their relaxation properties and sideband manifold (due to through space interactions with the substrate) will strongly depend on the substrate (i.e., paramagnetic or metallic electrodes) and measurements of their relaxation and resonance broadening can be used to estimate the thickness of the interphase (Figure 4).[42] Depending on the interphase composition or the polymeric components and their intimacy with the active electrode material they can be efficiently detected in NMR, for example through $^7$Li,

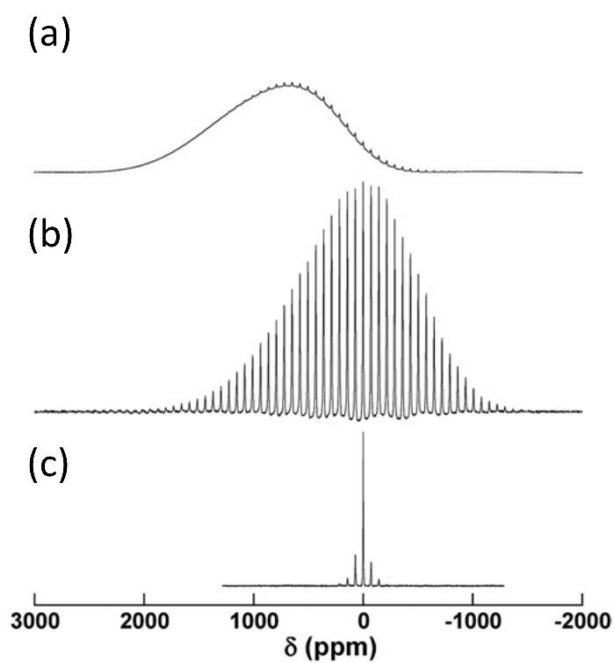

**Figure 4** $^7$Li MAS spectra of LiNi$_{0.5}$Mn$_{0.5}$O$_2$ stored in an ambient atmosphere for 2 months acquired with (a) Hahn echo showing mostly the broad resonance of paramagnetic Li from the bulk of the cathode where the paramagnetic broadening prevents resolution. The small sidebands on top are coming from Li$_2$CO$_3$ formed on the surface of the electrode in air. The diamagnetic surface species can be better resolved in (b) with single pulse excitation (as the broad component decays during the dead-time), demonstrating the significant breadth of sideband manifold which is not present in pure Li$_2$CO$_3$ (c). Longitudinal relaxation times were found to be 4 ms for the paramagnetic bulk Li, 200 s for pure Li$_2$CO$_3$ which shortens to 1 s when formed on the paramagnetic cathode substrate. Measurements were performed on 11.8 T (500 MHz) spectrometer with MAS of 14 kHz. Adapted from ref. 42 with permission.

$^1$H and $^{19}$F NMR for SEI phases such as LiOH, LiF and Teflon or PVDF binders. However, organic components such as polyethers and polycarbonates in the SEI can be practically invisible for NMR, especially if they are formed on paramagnetic substrates. Efficient detection of the organic components in the SEI layers on anodes such as graphite and Si was only practical with standard ssNMR through isotope enrichment of the solvent in the electrolyte.[43,44] Similarly, products formed in the SEI on RuO$_2$ or interphases formed in the carbon cathode in Li-O$_2$ cells could only be detected through $^{17}$O enrichment protocols.[45,46]

Thus, the fact that battery materials are attributed with very distinct magnetic resonance properties enables separation and assignment of chemical environments and components. This led to ssNMR developing into a powerful approach in the characterization of electrode and electrolyte materials. On the other hand, ssNMR is very limited when it comes to the study of native SEI layers or in the characterization of surface treatments which result in thin artificial SEI like layers. In this case, the combination of nuclei with low gyromagnetic ratio and/or low natural isotopic abundance found in thin surface layers can prevent the use of ssNMR. Nevertheless, interfaces and interphases play a pivotal role in the fate of battery materials, thus we must be able to characterize them in order to understand the function of battery cells. Furthermore, heterogeneity and disorder often prevent their characterization by techniques requiring long range order. Another major challenge associated with the study of SEI layers or non-passivating interphases forming in battery cells, is their chemical reactivity as they form under strongly reducing or oxidizing conditions. Thus, to gain reliable information into their composition they should be minimally perturbed by the method used for their study.

Compared to high-energy characterization tools, NMR is relatively nondestructive. Furthermore, its high chemical specificity and sensitivity to internuclear distances make it extremely informative in the study of interphases and interfaces. Thus, there is great interest in developing MAS-DNP approaches to unlock the potential of ssNMR and equip it with the needed sensitivity to probe interphases formed on electrodes and hidden interfaces between the different electrode and electrolyte components.

While interphases/faces are probably the main hurdle for ssNMR, employing DNP to gain sensitivity in the bulk will certainly benefit the study of many electrodes and solid electrolytes. By enabling the detection of nuclei with low gyromagnetic ratio and/or abundance such as $^{17}$O, $^{6}$Li, $^{29}$Si, $^{25}$Mg, $^{39}$K, and $^{43}$Ca through DNP, ssNMR can be used to gain structural insight, identify defects and their functional role and guide the design of new electrodes and solid electrolytes.

As will be discussed in the next section, the application of DNP to these materials is not always a straightforward extension of the current methodology. In the study of interphases, the metallic/paramagnetic nature of the substrate and changes in its magnetic resonance properties upon electrochemical cycling can have considerable effect on the DNP performance, in particular under cryogenic conditions. The chemical reactivity of the SEI and the inaccessibility of interphases and interfaces in composites, require tailoring the DNP approach to the materials and identifying endogenous sources of polarization.

In the following, we will describe how MAS-DNP has been employed to address several of these applications and how the unique properties of battery materials may lead to new DNP approaches.

## 4. DNP approaches for Batteries materials

DNP has seen great progress since its conception by Overhauser[47] and its realization by Carver and Slichter demonstrating the feasibility of DNP on lithium metal.[48] Through the development of instrumentation, efficient polarizing agents and understanding of the spin physics underlying the DNP process, high field MAS-DNP revolutionized the field of ssNMR. In the field of materials science, the implementation of DNP surface enhanced NMR spectroscopy (DNP-SENS), in which a solution of nitroxide biradicals is introduced into the material of interest, transformed the kind of materials and research questions that can be addressed by ssNMR spectroscopy.[49,50] The list of applications is impressive and was recently reviewed by Lafon et al.[9] In particular, the capabilities to characterize active sites grafted on the surface of heterogeneous catalysts[51,52] inspire further extension of the approach to study interphases in battery materials.

The basic ingredients of the DNP experiment are a source of electron spin polarization, namely a polarizing agent, microwave source that enables irradiation on the relevant transition and a MAS probe, commonly operated at cryogenic temperatures. The electron spin properties of the polarizing agents are crucial for the success of the experiment. We will now describe the three main DNP schemes used to increase the sensitivity in battery materials. For a full theoretical description of the different DNP mechanisms, we refer the reader to the recent reviews by Thankamony et al.[53] and Jardón-Álvarez et al.[54]

### 4.1. *Exogenous DNP: DNP-SENS*

The most common approach to boost the sensitivity of ssNMR to the surface of the material is via DNP-SENS. In this approach the sample is wetted or impregnated by a solution of nitroxide biradicals in a suitable solvent and cooled to cryogenic temperature (typically 100 K) in the DNP probe (Figure 5a). The prevalent formulation for materials applications is to add a solution of about 16 mM of TEKPol biradicals dissolved in tetracholoroethane (TCE). Radical concentration can of course be optimized and other organic solvents can also be employed.[55] Water based solution and water-soluble radicals may also be employed, though these are not suitable for studies of the native reactive SEI.

*Mechanism of polarization transfer*: Nitroxide biradicals are specifically designed to transfer polarization through the cross effect (CE). This mechanism requires a pair of coupled electrons with at least one of the electrons coupled through hyperfine interactions to a nuclear spin. Furthermore, the frequency difference between the electron spins has to match the Larmor frequency of the nucleus, $|\omega_{e1}-\omega_{e2}| = \omega_n$. Such conditions can be achieved by using radicals with g-anisotropy when the two radicals have different orientations with respect to the magnetic field. When such conditions are met, the polarization difference between the two electrons is transferred to the nucleus through the CE when microwave irradiation is applied to one of the single-quantum (SQ) electron spin transitions (Figure 5b). Cryogenic temperatures are required in order to freeze the solution in a glassy state thereby preventing radicals' aggregation as well as to slow down the electron longitudinal relaxation and achieve efficient saturation of the electron transitions. The polarization can be transferred from the radicals to the sample in two paths (Figure 5a): (i) Indirectly, where the $^1$H nuclei in the solvent are first polarized and the polarization spreads through $^1$H spin diffusion across the solvent (and sample if it contains protons). Polarization is then actively transferred to the heteronuclei at the surface of the sample through heteronuclear recoupling schemes (typically cross polarization or symmetry-based RF sequences). (ii) Directly, when the radicals have good affinity to the sample, such that they are found a few Angstroms away from its surface, the polarization can be transferred directly to the nuclei at the surface of the material.

The main advantage of DNP-SENS is that it is relatively general and often results in extremely high sensitivity gains (up to 4 orders of magnitude) to the surface and subsurface layers of the materials. Extending polarization to the bulk may also be possible in some cases as will be described later. Thus, it can be used efficiently to probe SEI layers and artificial coatings as will be described in the next section. Its main disadvantage is that the introduction of radicals and solvent may not be compatible with battery materials. Careful optimization must be performed to ensure that the sample stays intact (i.e., the surface layer is not washed with the solvent) and that there is no reactivity with the radicals. Development of sterically hindered radical species as was recently demonstrated in the study of reactive catalyst surfaces[56] will certainly benefit research on battery materials as well.

### *4.2. Endogenous DNP: Paramagnetic metal ions and defects*

An alternative approach to exogenous DNP is to use an inherent source of polarization that will minimally interfere with the chemistry of the material. These can be localized electrons as will be described in this part or delocalized as will be described in the next part. Localized sources of polarization can be introduced into the material of interest in the form of paramagnetic metal ion dopants with unpaired *d* or *f* electrons which can then be used to polarize the bulk and surface of the material (Figure 5c). This approach has emerged in the early days of DNP where paramagnetic impurities or dopants such as $Ce^{3+}$, $Fe^{3+}$ and $Cr^{3+}$ were employed to polarize single crystals of inorganic compounds at low fields and temperatures.[57–59] Corzilius et al. introduced this approach with modern high field MAS-DNP systems, demonstrating the suitability of $Gd^{3+}$ and $Mn^{2+}$ for polarizing biomolecules[60,61] as well as using $Cr^{3+}$ as a polarizing agent in a molecular crystal.[62] Our group has adapted the approach of metal ions DNP (MIDNP) for inorganic materials in general, and battery materials in particular.[63,64] Suitable polarizing agents are metal ions with half-filled shell configuration leading to isotropic resonances and minimal spin orbit couplings, and half integer electron spin resulting in a sharp EPR line. While in nitroxide radicals the EPR resonance is mostly broadened by g-anisotropy, for high spin metal ions the main source of broadening is the zero-field splitting (ZFS). The longitudinal relaxation times of metal ions are typically faster (in the µs scale) than nitroxides. Relaxation can be slowed down to some extent when the metal ions are doped in relatively symmetric environments leading to minimal ZFS[65,66] which requires minimal perturbance to the crystal lattice.

*Mechanism of polarization transfer*: The most common polarization transfer mechanism for high spin metal ions is the solid effect (SE). The SE mechanism involves one nucleus and one electron that are coupled by hyperfine interactions, resulting in a mixing of their spin states. This allows transfer of polarization when irradiating the normally forbidden electron-nuclear transitions (double or zero quantum, DQ and ZQ respectively). To satisfy this condition, the microwaves frequency needs to be equal to the sum or difference of the electron and nuclear Larmor frequencies $\omega_{\mu w} = \omega_e \pm \omega_n$ (Figure 5d). For high spin metal ions this transfer is most efficient for the central transition (connecting the electron spin states $|1/2>$ and $|-1/2>$) since it is minimally broadened by the ZFS interaction. Polarization transfer through the lattice is then achieved either through spin diffusion (in case of polarizing relatively sensitive and abundant nuclei such as $^7$Li or $^{31}$P) and/or directly to all nuclei when the paramagnetic metal ion is the main driving source of nuclear relaxation.[67] The typical concentration range optimal for MIDNP is 5-80 mM of dopant. Some optimization is required to obtain optimal enhancement, with 35 mM generally providing sufficient sensitivity. While in principle possible, CE is less likely to be encountered as the polarization mechanism with metal ions. For CE to occur under MAS, the coupled electron spins have to have different orientations so that their frequency is separated by the nuclear Larmor frequency. In crystalline solids this is only possible if two metal ions dope different crystallographic sites (giving rise to different ZFS tensors) or through CE matching of the central and satellites transitions. The latter was theoretically discussed by Corzilius et al.[68] but yet to be observed experimentally. Furthermore, under MAS the CE is a consequence of a cascade of rotor events during which the electrons have to maintain their polarization difference. This requires relatively long electron relaxation times which are not commonly observed with metal ions. Nevertheless, as the range of materials tested with the MIDNP approach increases, new avenues for polarization transfer may be identified.

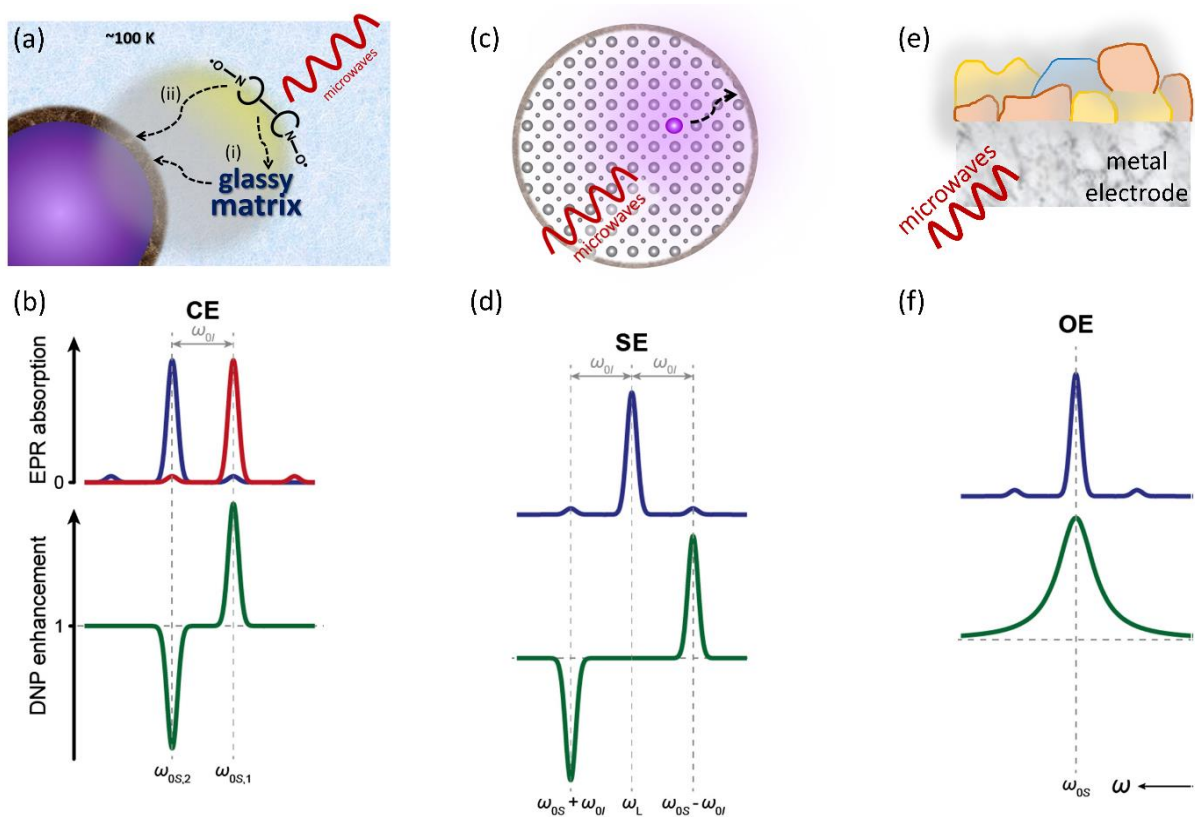

**Figure 5** The different approaches for DNP on battery electrodes: (a) DNP-SENS: A solution of nitroxide biradicals is wetting the sample of interest (purple ball) and frozen in the MAS probe at cryogenic temperatures. Polarization from the electron spins can be transferred to the sample through microwaves irradiation (i) indirectly, where the polarization is first transferred to the $^1$H nuclei in the frozen solvent and then by heteronuclear recoupling sequences to the surface of the sample, or (ii) directly, where the nuclei at the surface of the sample are directly polarized by the radicals. (c) Endogenous DNP from paramagnetic metal ion dopants introduced into the bulk: Polarization from high spin metal ions is transferred to nuclei in the bulk of the materials or its surface, typically via direct SE polarization transfer. (e) OE-DNP from conduction electrons: irradiation of the EPR transition of the conduction electrons results in polarization transfer to the metal nuclei through cross relaxation processes. Polarization can also be selectively transferred to the surface either directly or through spin diffusion from the metal. The three different mechanisms can in principle be distinguished by their DNP sweep profiles: (b) Irradiation on one of the SQ electron transitions ($\omega_{0S,1}$ or $\omega_{0S,2}$) of two coupled electrons separated by the nuclear Larmor frequency, $\omega_{0I}$, where at least one coupled electron to the nucleus leads to CE DNP. (d) Irradiation on a ZQ or DQ transitions at $\omega_{0S}-\omega_{0I}$ or $\omega_{0S}+\omega_{0I}$, respectively, will lead to SE DNP in the case of a hyperfine coupled nuclear-electron pair. (f) Irradiation on the EPR transition of conduction electrons at $\omega_{0S}$ along with electron-nuclear cross relaxation leads to OE DNP. Panels b, d and f were adapted ref. 53 with permission.

Cryogenic temperatures are still commonly employed, again in order to increase electron relaxation times and enable efficient saturation of the ZQ/DQ transitions. Nevertheless, it was recently demonstrated by Hope et al. that highly symmetric environments may lead to relatively long electron relaxation times which enable efficient polarization even at room temperature and above.[69]

The main advantage of MIDNP is that it provides high sensitivity in the bulk of inorganic materials which is often difficult or impossible to achieve with exogenous polarization sources. Polarization can further extend to the surface of the material as well as interphases deposited on it. Depending on the composition and relaxation properties of the interphases, the polarization can be transferred directly from the metal ion dopants, through spin diffusion from the bulk, or indirectly via heteronuclear polarization transfer. We note that this approach is relatively new and there are many open questions and room for optimizing polarization transfer to interphases. Another major advantage of the approach is that it minimally interferes with the chemistry of the materials, thus it is compatible with probing reactive interfaces and interphases.

The main disadvantages of MIDNP are that it is not a general approach; the polarization source and conditions for optimal DNP have to be optimized depending on the sample. The dopant can be chosen either as a structural spy, in that case it should perfectly fit in the structure in terms of charge and ionic radii. Or it can be chosen to improve the materials properties, for example to increase the electrochemical performance either through increased ionic conductivity or structural stability. Finally, the technical constraints of current commercial MAS-DNP systems, which have limited field sweep capabilities and fixed microwave frequency sources, place some constrains on the choice of metal ions.

Endogenous sources of polarization that are inherent in the materials are not very common. A relevant example is the use of localized defects in carbonaceous materials[70] and the seminal work by Wind et al. who utilized multiple DNP mechanisms from the different polarization sources in coals.[71]

### *4.3.   Endogenous DNP: mobile electrons*

When the electron spins are mobile it leads to averaging of the through space dipolar interactions making SE and CE non-feasible. This is the case for radicals in solution and for mobile electrons in solids found in the conduction bands of metals or in delocalized bonds in conjugated systems. In these cases, electron-nuclear cross relaxation processes can give rise to polarization transfer as described below.

*Mechanism of polarization transfer*: The dominating mechanism in systems with mobile electrons, such as liquids or metals, is the OE mechanism. As mentioned above, the first DNP experiment was performed by T.R. Carver and C.P. Slichter in the 1950s, proving the mechanism predicted by Overhauser by utilizing the conduction electrons in metallic lithium.[48] The OE mechanism involves the irradiation of the SQ EPR transition, resulting in saturation and equalization of the electron spin state populations. This can result in polarization building up across the NMR transitions in the presence of favorable electron-nuclear cross relaxation processes (Figure 5f). Efficient cross relaxation requires significant electron-nuclear interactions that are fluctuating with a spectral density component at approximately the electron Larmor frequency. The main electron-nuclear interactions are through space (dipolar) and through bonds (Fermi) which will give rise to DQ and ZQ (dipolar) or ZQ (Fermi) cross relaxation processes. As the two pathways result in opposite sign for the nuclear hyperpolarization, OE DNP is most efficient when one of the pathways dominates.

In liquids, the motion of radicals leads to modulation of the through space dipolar interactions with solvent $^1$H molecules and through bonds couplings to $^{13}$C via collisions with the solvent.[72,73] Of course, the efficiency of OE in liquids strongly depends on the timescale of these motions (thus on the viscosity of the liquid) and the magnetic field of the measurement.

This approach was not examined in depth in the context of battery materials, with only one example where electrochemically active radicals were used to polarize the electrolyte.[74] Extension of this approach further to study low concentration molecular species in liquid electrolytes (for example to follow electrolyte degradation) has not been explored yet.

In metals, the high mobility of electrons in the conduction band provides an efficient source of ZQ cross relaxation through the Fermi contact shift. Thus, in principle, one can envision that using the conduction/delocalized electrons in the electrode materials (either of the active materials or the conductive carbon additive) to polarize the nuclei in the bulk or at the surface of the electrodes would be an excellent path for DNP in battery materials (Figure 5e). This approach has been recently demonstrated by Hope et al. who used the conduction electrons in lithium dendrites to polarize the metal itself as well as its SEI.[75] Remarkably, up to 10-fold signal enhancement was also achievable at room temperature for the SEI.

The main advantage of this approach is that it offers a non-invasive route for DNP with no modifications to the material and may open the path for DNP at room temperature and possibly full battery cells. As there is only one reported case of OE-DNP from metals at high field there are many open questions such as the mechanism of polarization transfer from the metal to nearby phases. While for lithium it is very likely that spin diffusion is the mechanism spreading the polarization from the metal and across the SEI, heteronuclear polarization transfer is less understood. Extension of this approach to other metals or systems with delocalized electrons is a very interesting avenue to explore.

### *4.4.  Considerations when choosing the DNP approach*

The magnetic resonance properties of battery materials vary significantly between different components and may limit the applicability of DNP or give rise to specific DNP paths. Here we provide general guidelines based on our experience and the examples reported in the literature which will be described in more detail in the next section. A summary is provided in Table 1.

DNP SENS is the most efficient approach for gaining high sensitivity to the electrodes surface, the SEI formed on them as well as thin coatings used for surface passivation. However, when employing this approach, one has to consider possible reactivity with the sample, which can be tested by following the EPR signal of nitroxide biradicals from the time they are added to the sample. Furthermore, the generality and efficiency of this approach to interphases deposited on paramagnetic electrodes is not clear. Paramagnetic electrodes lead to significant shortening of the relaxation time of nearby nuclei and electron spins which may limit the sensitivity gain from DNP. Importantly, the magnetic properties of paramagnetic materials change significantly with temperature. With cooling, paramagnetic materials often have higher magnetic susceptibility and many cathode materials may become magnetically ordered when cooling to 100 K and below. For example, $LiNi_{0.5}Mn_{1.5}O_4$ becomes ferromagnetic below 100 K[76] while $LiFePO_4$ is antiferromagnetic below about 50 K with higher transition temperatures reported depending on the synthesis conditions.[77] This may again change the relaxation properties of the sample, lead to significant spectral broadening as well as pose technical difficulties when the sample is strongly magnetic. Thus, we expect the DNP-SENS approach to be suitable to study surface and subsurface layers of diamagnetic or weakly paramagnetic substrates as listed in Table 1.

| **DNP approach** | **materials** | **type of information** | **comments** |
|---|---|---|---|
| DNP-SENS | Li anodes: graphite, graphene, Si, titanates | SEI composition coatings composition | Chemical compatibility of radical solution should be verified |
| | Na anodes: titanates, phosphates, hard carbon | SEI composition coatings composition | |
| | Weakly paramagnetic cathodes: ruthanates, $LiCoO_2$* | | *Diamagnetic in its lithiated state |
| MIDNP | Diamagnetic anodes: titanates, phosphates | Local order and transformation in the bulk Interphases | Studies of uncycled materials or following full cycles to avoid presence of other paramagnetic ions. Electrochemical activity of dopant should be tested. |
| | Solid electrolytes | Structural insight, defects formation, vacancies ordering | |
| | Composite electrolytes | Structural insight, polymer-ceramic interfaces | |
| OE – metals | Metal anodes | SEI composition | Extension to conductive non-metallic materials to be tested. |
| OE – liquids | Liquid electrolytes | Degradation of electrolyte | OE through electrochemical active radicals or stable radicals in solution. |
| SE/CE DNP | Liquid electrolytes in glassy state | Structural insight into solvation shells | |

**Table 1** Summary of the current DNP approaches and the type of battery materials and insight they can be used to study.

The same limitations concerning paramagnetic electrodes are even more significant in the case of MIDNP as the presence of many paramagnetic metal ions will result in significant shortening of the dopant relaxation time through strong electron dipolar and exchange interactions. Thus, MIDNP is most suitable to study interphases and the bulk of diamagnetic substrates that are doped with the polarizing metal ion (see Table 1), as well as the bulk and interfaces in solid and composite electrolytes (where the dopant is introduced to the ceramic particles). We also note that electrochemical cycling may change the magnetic properties of the materials (so some materials may only be studied in their charged/discharged states). Furthermore, changes in the oxidation state or EPR properties of the dopant should be considered and avoided if not reversible.

OE-DNP in the solid state can of course be employed to study the SEI on metal anodes. Extension to other conductive but non-metallic materials remains to be further explored. OE-DNP in liquids can potentially be used to probe liquid electrolytes and their composition. The possibility to transfer polarization to electrode interface through collisions with radicals should be examined. Clearly liquid state OE-DNP for battery electrolytes has to be developed to test its viability.

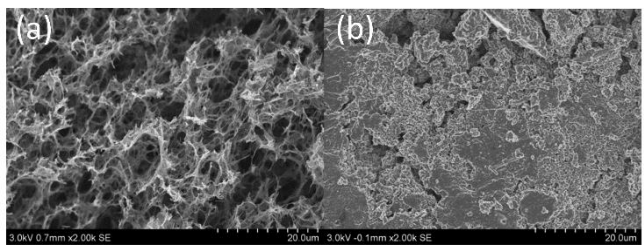
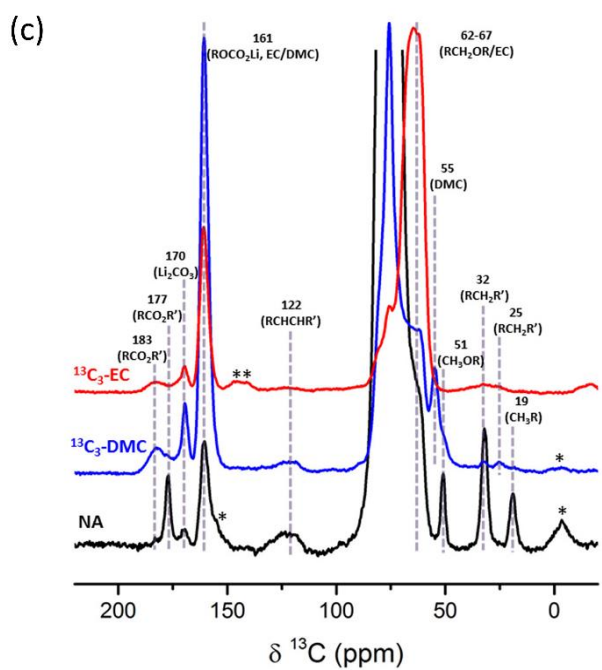

**Figure 6** Scanning electron microscopy images acquired (a) before and (b) after lithiation of rGO electrode. (c) DNP-SENS $^1$H–$^{13}$C CP spectra of the rGO electrode cycled in natural abundance (NA) electrolyte (black) and in electrolyte with $^{13}$C$_3$–Dimethyl carbonate (DMC, blue) and $^{13}$C$_3$–Ethylene carbonate (EC, red). All spectra were acquired with a contact time of 3 ms, relaxation delay of 7 s, and microwave irradiation at 9.4 T, MAS of 10 kHz and temperature of approximately 100 K. Figure was adapted from ref. 80 with permission.

Lastly, we note that while conductive materials may be a source for DNP, they can also lead to significant sample heating and poor DNP enhancement due to microwave absorption. In our work we have found that conductive carbon additives, such as carbon black, are extremely deleterious to the DNP process, lowering the enhancement of solvent signals in DNP SENS by up to 2 orders of magnitude, depending on the degree of carbon conductivity and its amount in the sample.[78] Thus, for DNP investigations, carbon-free electrode formulations can be employed when possible, or the amount of carbon should be minimal.[79] Furthermore, for decreasing the heating effect the samples can be diluted with KBr for example.

## 5. Applications of DNP to Battery Materials

### *5.1. Sensitive detection of interphases*

Enabling high sensitivity detection of interphases deposited on the surface of electrodes or thin coatings used as artificial SEI by ssNMR is probably the most important contribution of DNP to the study of battery materials to date. The first example employing DNP-SENS to probe the SEI with high sensitivity, was performed on reduced graphene oxide (rGO) anodes.[80] In this case rGO was chosen to test the approach as it has high surface area and significant irreversible capacity on its first cycle indicating extensive formation of SEI on its surface. MAS-DNP enabled acquiring $^1$H–$^{13}$C CP spectra within a few hours without isotope enrichment. Spectra acquired from cycled rGO revealed the fingerprints of the organic and inorganic components of the surface layer (Figure 6). Surprisingly, when applying this approach to rGO anodes that were cycled in $^{13}$C enriched electrolyte and thus incorporated a $^{13}$C enriched SEI, no enhancement could be obtained. This observation was rationalized by the formation of very thick SEI trapped within the pores of the rGO framework which prevented efficient spin diffusion from the surface to the "bulk" of the SEI. When the majority of the SEI is $^{13}$C rich, its signal dominates over the outer SEI layers even though those are enhanced by DNP. Heating due to the conductive rGO substrate could also lead to poor DNP performance for the SEI layers in contact with the rGO.

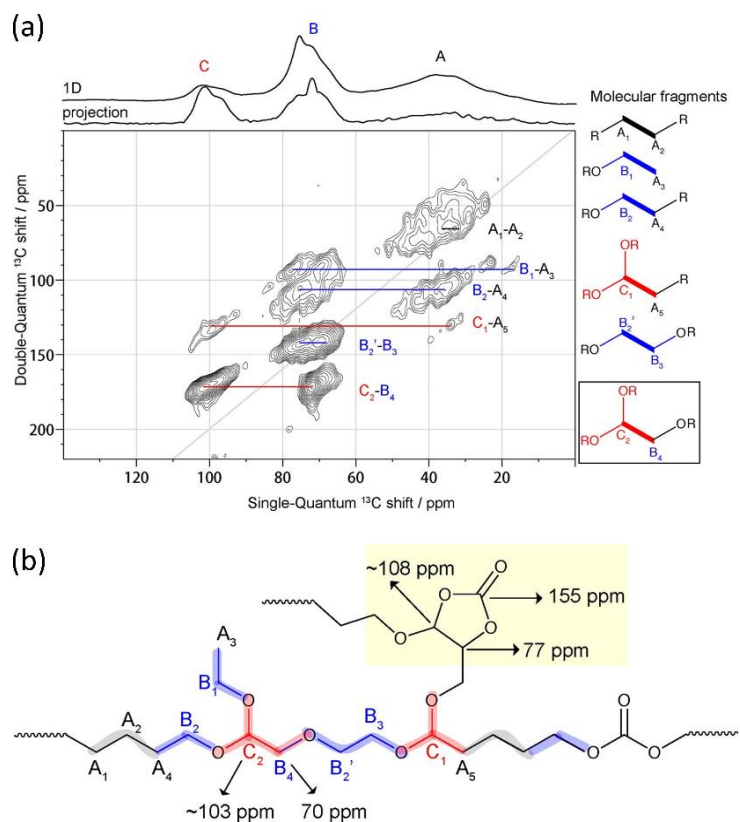

**Figure 7** (a) DNP-SENS enabling 2D DQ–SQ $^{13}$C–$^{13}$C POST-C7 dipolar correlation spectrum of Si nanowires cycled in fluoroethylene carbonate (FEC) EC + 5% $^{13}$C$_3$ FEC for 100 cycles. Top: The 1D $^1$H–$^{13}$C CP NMR and the total projection in the SQ dimension. The corresponding molecular fragments are shown in the right panel. The branching fragments are colored in red; the fragments containing ethylene oxide carbons are colored in blue, and the alkyl chains in black. The spectrum was recorded at 14.1 T (600 MHz), MAS of 12.5 kHz, 105 K and took ~9 h to acquire. (b) Possible molecular fragments observed in the FEC/vinyl carbonate decomposition products. Adapted from ref. 84 with permission.

A promising solution to overcome the formation of an unstable spontaneous SEI layer on the electrode surface is through the incorporation of electrolyte additives.[81,82] Such additives are chosen as the energy of their 'lowest unoccupied molecular orbital' (LUMO) is lower in comparison with the electrolyte main constituents making them more susceptible to reduction. Beneficial additives should be reduced prior to the electrolyte, thereby passivating the composite electrode and suppressing the degradation of the electrolyte.[81,83] Jin et al. studied the surface layer formed on silicon nanowires in the presence of fluoroethylene carbonate (FEC) and vinylene carbonate (VC) additives.[84,85] The enhancement in signal to noise ratio (SNR) together with the reduction of experiment time achieved through DNP-SENS enabled acquiring 2D dipolar correlations (Figure 7) which allowed them to establish the cross linked framework of the organic species in the SEI. Combination of DNP enhanced direct excitation $^{29}$Si experiments with $^1$H-$^{29}$Si CP experiments disclosed the interfacial silicon groups and further proved that the silicon nanowires are bonded to organic species and not fully covered by an inorganic interface as previously reported. The improvement in electrochemical performance of the anode was explained by this discovery together with the coverage uniformity of the additive derived SEI of the silicon nanowires.

One of the biggest drawbacks of the requirement of cryogenic temperatures in DNP is the limitation it places on testing paramagnetic materials, namely metal oxide cathodes. Nevertheless, the surface layer formed during cycling on the cathode material $Li_2RuO_3$ was identified by DNP-SENS due to the relatively reduced magnetic susceptibility of the cathode.[86] $^1$H-$^{13}$C CP DNP and room temperature ssNMR measurements allowed a comparison of the electrolyte decomposition species deposited on the surface of the cathode at the charged and discharged states (Figure 8). Surprisingly it was found that the interphases deposited on the cathode are very similar in their composition to those found in the SEI on anodes, containing mostly PEO type species, lithium acetate, carbonate and fluoride. Furthermore, the composition of the interphases was distinct on charge vs. discharged state, which was explained by migration of surface species from anode to cathode. The chemical composition of the interphases as revealed by DNP-NMR was correlated with the electrochemical impedance response of the cathodes, leading to the conclusion that the phases are blocking interfacial lithium transport.

**Figure 8** Two-dimensional (2D) $^1$H-$^{13}$C heteronuclear correlation enabled by DNP-SENS of $Li_2RuO_3$ cycled against Li in LP30 and disassembled at the end of the 1st charge/discharge cycle at 2.0 V. (b-c) $^1$H-$^{13}$C CP MAS ssNMR and DNP-NMR spectra of the phases formed on $Li_2RuO_3$ at different states of charge and cycle number. The cells shown in (b) were disassembled in the charged state at 4.6 V after a single charge (blue, bottom left, DNP-NMR at 14.1 T at 100 K) and 100 cycles (black, top left, ssNMR at 14.1 T, room temperature). The cells shown in (c) were disassembled in the discharged state at 2.0 V after one charge/discharge cycle (blue, bottom right, DNP-NMR at 9.4 T at 92 K) and 27 cycles (black, top right, DNP-NMR at 9.4 T at 92 K). All samples were cycled in LP30 against Li metal. The grey rectangle in (c) is used to label peaks at 148 ppm and 160 ppm in the ROCO$_2$R′ region. Asterix denote spinning sidebands. Adapted from ref. 86 with permission.

An efficient approach to control the properties of the surface layer of the cathode and prevent electrolyte degradation is by creating an artificial SEI on the cathode. Such interphase can stabilize the surface, minimize side reactions and enhance favorable lithium transport. An example for such an efficient artificial SEI layer is an alkylated lithium silicate coating formed through a novel molecular layer deposition process. High energy cathodes based on a lithium rich layered Ni, Mn, and Co oxide, showed remarkable electrochemical performance in comparison to the uncoated cathode material.[87] As the coating thickness was limited to only 2-5 nm it was impossible to probe it by ssNMR. Thus, DNP was essential to gain detailed insight into the coating's chemical, structural and functional properties. In this case, due to the strong paramagnetic nature of the cathode and in order to avoid ambiguity with $^7$Li resonances from the bulk, DNP studies were performed on a model $TiO_2$ electrode that went through the same surface treatment as the cathode.[88] Exogenous DNP-SENS provided high $^1$H polarization which was then transferred via $^1$H-$^{29}$Si, $^1$H-$^{13}$C, and $^1$H-$^7$Li CP to the surface species. 1D spectra revealed the deposition of amorphous silica terminated with siloxanes and alkylated (tert-butyl and methyl) silicon environments. Lithium was found to be dispersed throughout the coating layer. The high sensitivity also enabled acquiring $^1$H-$^{29}$Si-{$^7$Li} CP rotational echo double resonance (REDOR) measurements which were used to determine phase segregation between Si and Li species. To gain structural insight into this amorphous coating, endogenous MIDNP was employed. In this case $Fe^{3+}$ ions were doped in the $TiO_2$ substrate. Comparison of $^{29}$Si and $^7$Li spectra obtained with direct polarization from endogenous ($Fe^{3+}$) and exogenous (nitroxides) sources (Figure 9), was used to construct a 3D model for the coating. This was based on the two approaches, DNP-SENS and MIDNP, which provide sensitivity to the outer coating-electrolyte and inner coating-electrode interface, respectively. This approach is expected to be a powerful structural tool for other thin coatings and SEI layers.

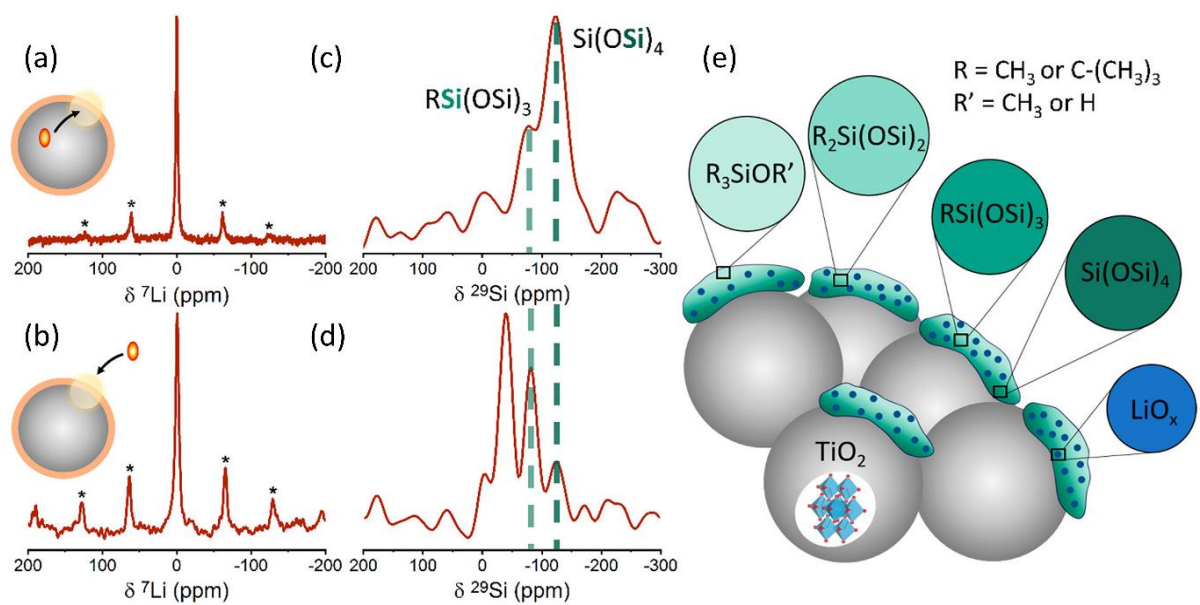

**Figure 9** Top spectra: direct polarization via endogenous DNP from $Fe^{3+}$ (inset: polarization source represented as red ellipse) to (a) $^7$Li nuclei and (c) $^{29}$Si nuclei using CPMG detection. Bottom spectra: direct polarization via exogenous DNP from the TEKPol (inset) to the (b) $^7$Li nuclei and (d) $^{29}$Si nuclei by using CPMG detection. Experiments were performed at 100 K with 10 kHz spinning speed. Monoalkylated silica and silica groups are marked with light green and dark green dotted lines, respectively. (e) A structural model of the $Li_xSi_yO_z$ coating layer showing the various silicon environments in different shades of green. Uniformly distributed $LiO_x$ is shown in blue. Adapted from ref. 88 with permission.

Metallic anodes offer a unique way to probe the SEI layers formed on them by selectively enhancing resonances in the SEI via OE DNP. Hope et al. demonstrated that this approach can be used to probe the surface layer formed on lithium dendrites at room temperature (Figure 10).[75] Only SEI species in close proximity to the lithium metal surface were hyperpolarized due to OE, leading to selective enhancement (as can be seen in Figure 10(b))

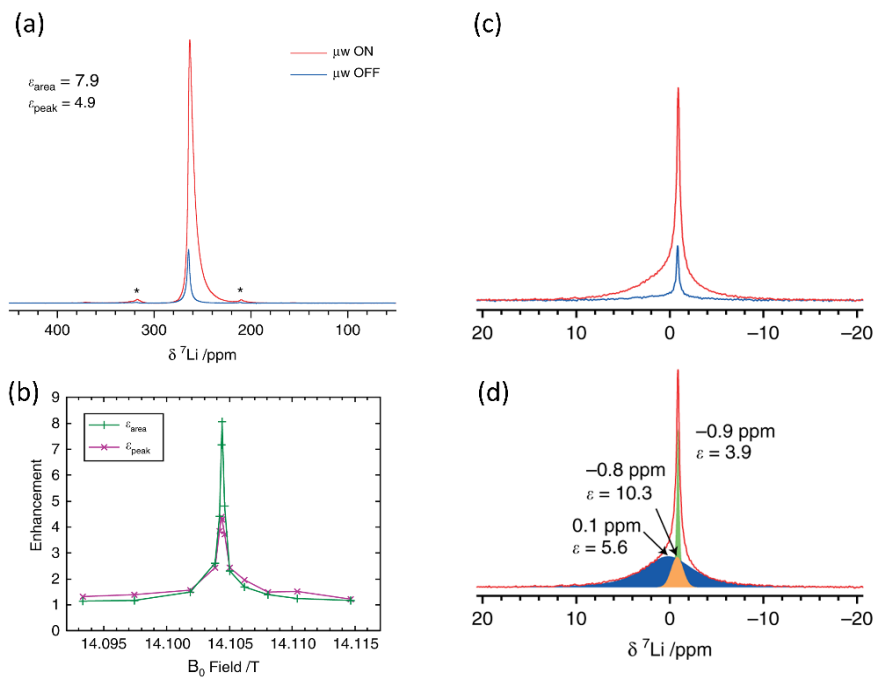

**Figure 10** (a) $^7$Li NMR spectrum of microstructural lithium metal, with and without 15.6 W of microwave irradiation at 395.29 GHz (μw ON/OFF), recorded at 14.1045 T, 12.5 kHz MAS and a sample temperature of ~300 K. Spinning sidebands are marked with an asterix. (b) The enhancement of the integrated intensity and peak intensity as a function of the $B_0$ field, measured at 100 K. (c) $^7$Li NMR spectra of lithium SEI produced by cycling with LP30 + FEC electrolyte, recorded with and without 15.6 W of microwave irradiation (μw ON/OFF). (d) Deconvolutions of the μw ON spectra in (c) recorded at 12.5 kHz MAS, 14.1 T and room temperature. Adapted from ref. 75 with permission.

and determination of the SEI structural arrangement. The feasibility of performing static OE-DNP experiments at room temperature suggests there is potential in developing this approach for in-situ battery set ups. In this case, OE can be utilized to enhance the SEI signals at electrochemical conditions and temperature relevant for their function.

*5.2. High sensitivity in the bulk of electrodes and electrolytes*

The use of exogenous DNP results in the highest boost in signal enhancement. However, since the polarization spreads from the solvent nuclei to the particles, the sensitivity is typically limited to the surface and subsurface layers. This of course depends on the material's composition and the affinity of the solvent and radicals to the sample. Björgvinsdótir et al. demonstrated that polarization from nitroxides biradicals can also propagate beyond the surface when efficient spin diffusion is present.[89] The effect was shown on various oxides[64] as well as on the anode material, $Li_4Ti_5O_{12}$ (LTO).[89] Comparison of the spectra acquired from $^7Li$ direct polarization and $^1H$-$^7Li$ CP experiments revealed different chemical sites due to the different polarization pathways.

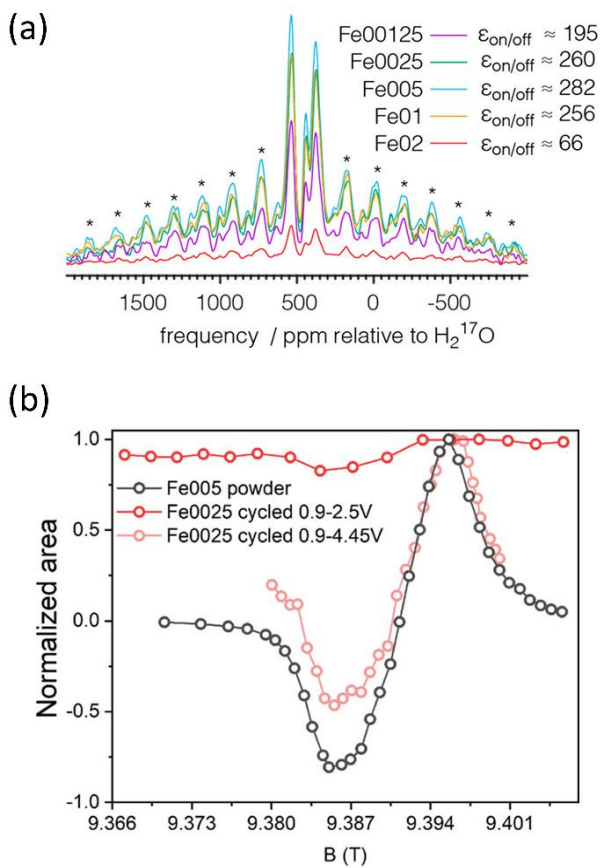

**Figure 11** (a) Natural abundance $^{17}O$ DNP NMR spectra of $Fe^{3+}$-doped $Li_4Ti_5O_{12}$. Spectra were obtained at 9.4 T, 100 K, 10 kHz MAS with 16 scans (except Fe02LTO, which was obtained with 128 scans. The spectrum in the figure was scaled accordingly). (b) $^7Li$ DNP sweep profiles acquired with a build-up time of 10 s for $Fe^{3+}$-doped $Li_4Ti_5O_{12}$ before and after galvanostatic cycling. Only when the sample was cycled with a broad voltage range, $Fe^{3+}$ was formed again and enabled DNP enhancement. Adapted ref. 67 and 79 with permission.

MIDNP offers an efficient way to gain sensitivity across the bulk of electrode materials, independent of the particle size and in the absence of efficient spin diffusion. The first example of this approach to an anode material was with static DNP where polarization from $Mn^{2+}$ dopants was transferred to surrounding $^7Li$ nuclei in LTO, leading to a 14-fold enhancement of the $^7Li$ resonance.[63] Higher enhancement was demonstrated with MAS-DNP,[64] leading to 24 and 140 fold increase in signal intensity for $^{7,6}Li$ respectively and remarkably enabling the detection of $^{17}O$ at natural isotopic abundance of only 0.038%. The enhancement was investigated as a function of the dopant concentration, revealing that at low concentrations the lithium spectra are uniformly enhanced (i.e. the spectra with and without DNP were identical in line shape) but with increasing dopant concentration broader environments were better enhanced. As $Mn^{2+}$ was found to replace lithium on tetrahedral sites in the LTO spinel lattice it leads to reduced rate performance of the anode with increasing concentration.[79] $Fe^{3+}$ in LTO was found to be a much more efficient polarizing agent for MIDNP, with $^{17}O$ signal enhancements approaching 300 (Figure 11a).[67] As a dopant and substituent up to a concentration of 2% (mole) it was found to provide improved capacity retention in LTO.[79] Unlike $Mn^{2+}$, $Fe^{3+}$ is not electrochemically inert. Harchol et al. found that the dopant participates in the redox process (reducing to $Fe^{2+}$ upon lithiation of LTO) which inevitably led to loss of DNP activity as $Fe^{2+}$ is EPR silent. However, reactivation of the Fe dopants was possible by increasing the electrochemical voltage window, as was confirmed by the reappearance of the $^7Li$ DNP sweep profile following electrochemical cycling (Figure 11b). In this study, the importance of optimizing the electrode formulation for DNP investigation was also explored, demonstrating that carbon-free electrode formulations give rise to higher DNP enhancements with minimal effect on the electrochemical performance.

The application of the MIDNP approach is mostly suitable for the study of bulk properties of diamagnetic electrode material, such as alkali titanates and titanium phosphate anodes. Another possible application that was not explored yet is the implementation of MIDNP in solid

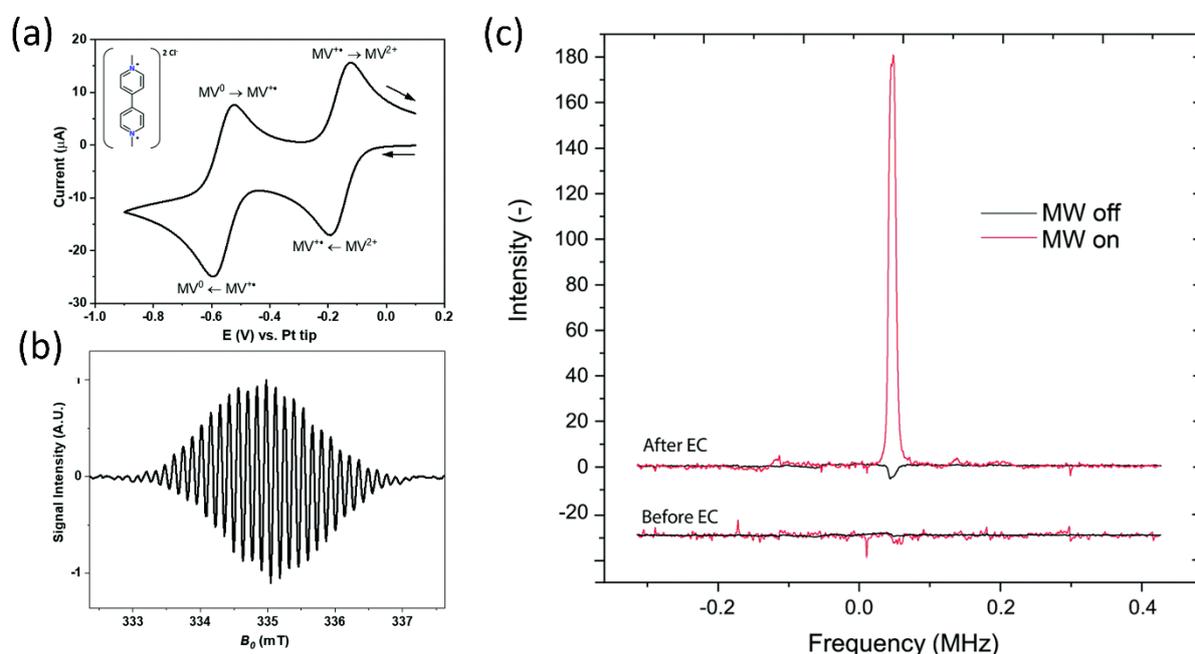

**Figure 12** (a) Cyclic voltammogram of a 1 mM solution of $MV^{2+}$ in acetonitrile with 200 mM tetrabutylammonium perchlorate, TBAP. The redox reactions are indicated on the voltammogram. Inset: The chemical structure of $MV^{2+}$. (b) The EPR spectrum of ca. 100 μM of $MV^{+\cdot}$ in acetonitrile with 200 mM TBAP. (C) $^1H$ NMR spectra at 14 MHz. Spectra before and after the electrochemical (EC) process. The enhancement calculated from the magnitude of the different spectra is 31. The "Before EC" spectrum is offset by −30 arbitrary units. Adapted from ref. 74 with permission.

electrolytes. In this case, sensitivity from MIDNP can be used to gain structural understanding of the crystalline ceramic phases (for Li, Na, and Zn batteries). We have recently utilized MIDNP from $Gd^{3+}$ to rule out oxygen vacancies clustering in Y doped $CeO_2$.[90] Vacancies clustering has an important role in the ionic conductivity of oxygen ion conductors for solid oxide fuel cells, and we expect the approach to be insightful also for cationic conductors. Furthermore, MIDNP offers a way to gain selectivity and sensitivity in probing interfaces which may not be accessible for exogenous sources of polarization, for example interfaces formed in composite electrolytes.

Liquid electrolytes can be probed at room temperature with high sensitivity through OE-DNP. An original approach, utilizing electrochemically active organic radicals was recently given by Tamski et al. with redox mediator methyl-viologen dication ($MV^{2+}$) which produces a radical during electrochemical cycling.[74] $^1$H OE-DNP resulted in solvent enhancement of 30 from electrochemically generated radicals (Figure 12). This approach offers a new path to in-situ characterization of electrochemical cells, utilizing radicals which can be electrochemically activated.

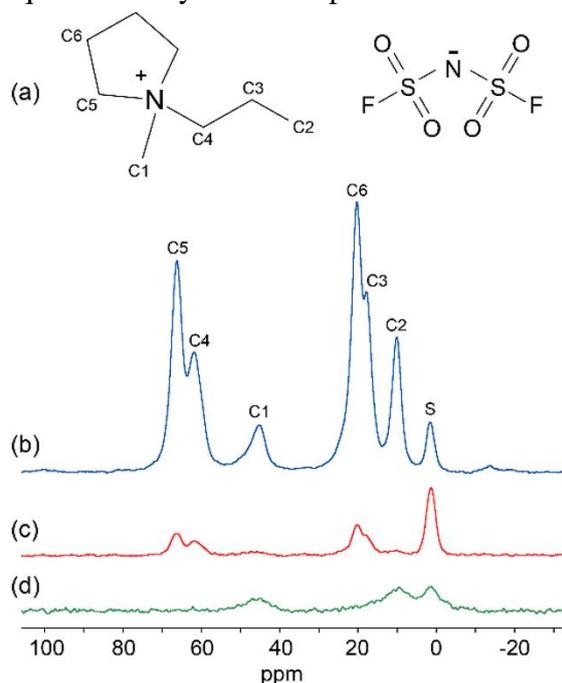

**Figure 13** (a) Molecular structure of the $C_3$mpyrFSI ionic liquid with the six inequivalent carbon sites labeled. $^{13}$C CP MAS spectra obtained from the frozen $C_3$mpyrFSI-$^6$LiFSI-TEKPol solution are shown with (b) DNP enhancement and $^1$H decoupling, (c) no DNP enhancement and $^1$H decoupling, and (d) DNP enhancement but no $^1$H decoupling. All spectra were acquired at 9.4 T and 8 kHz MAS at ~92 K. The peak assignment is shown in panel b, where S denotes the signal from the silicone plug. Adapted from ref. 91 with permission.

Finally, the structure of liquid electrolytes can also be studied in their frozen state. Sani et al. studied an ionic liquid (IL) electrolyte at 92 K where DNP was enabled by mixing the electrolyte with TEKPol.[91] This enabled rapid detection of low abundance and low sensitivity nuclei such as $^6$Li and $^{13}$C and was used to characterize the nanoscale structure of an IL containing lithium bis(fluorosulfonyl)imide (LiFSI). $^1$H-$^{13}$C CP, acquired under DNP conditions, revealed new carbon surface species on the IL that were otherwise undetected (Figure 13). 3D spatial probability plots were constructed from $^{13}$C-{$^6$Li} REDOR experiments and further supported with molecular dynamic simulations to find the proximity of lithium and carbon nuclei, shedding light on the structure of the IL solution.

## *6. Summary and Outlook*

In this trends article we discussed the current capabilities, limitations and possible developments associated with the application of DNP to battery materials. As batteries play a pivotal role in the transition to utilization of sustainable energy resources there is great interest in developing new and improved battery systems. Implementation of new materials and chemistries with reversible charge storage capabilities requires in depth characterization of the materials' properties and functionality. In this regard, ssNMR spectroscopy is increasingly utilized, with well-developed methodology for probing the bulk of lithium and sodium-based electrode materials. Following the widespread use of DNP in materials science applications such as heterogeneous catalysis, it can be expected that DNP approaches can also be incorporated in the study of battery materials. In particular, recent examples clearly demonstrate how DNP equips ssNMR with the needed sensitivity to probe one of the most important components in the battery - the electrode-electrolyte interface. Implementation of different approaches for DNP, namely DNP-SENS, Overhauser DNP from conduction electrons in metals and endogenous DNP based on paramagnetic metal ions, all provide an efficient route to probe naturally or artificially forming SEI layers. DNP-SENS and MIDNP were also shown to provide significant bulk sensitivity in different cases. This can aid characterization of new solid electrolytes and electrodes with ssNMR by enabling detection of otherwise challenging nuclei.

Based on these first investigations there is a lot of room for development of new applications for DNP as well as new approaches which will increase its capabilities. In this section we would like to survey our perspective on these developments.

In particular, one family of materials that is less amenable to DNP, especially at cryogenic temperatures, are strongly paramagnetic cathodes. As mentioned above the effect of paramagnetic substrates on DNP-SENS has not been explored yet. As each material may have its own unique dependence on temperature in terms of its magnetic response this would require comparing different types of magnetic phase transitions and their effect on the ability to polarize diamagnetic interphases. This family of materials is also not compatible with MIDNP and alternative routes for utilizing the paramagnetic metal ions could possibly be identified in order to gain sensitivity from endogenous polarization sources.

Paramagnetic metal ions are relatively new polarization sources in MAS-DNP. Their main advantage is that they can be thought of as structural spies. As such we expect to see more studies where MIDNP is used to elucidate structural properties of new solid electrolytes or anodes based on diamagnetic oxides and phosphates for the growing family of alkali and earth alkaline storage materials. The conditions for polarizing interphases through the MIDNP approach should be investigated and the extent of polarization transfer should be explored in different interphases depending on their composition. The combination with DNP-SENS is extremely promising for elucidating the mosaic structure of the SEI as well as to guide the rational design of artificial SEI layers.

The utilization of Overhauser DNP from radicals in liquids can be developed as an excellent tool with which to study the electrolyte composition and structure. For example, this approach can be developed to gain insight into the electrolyte solvation shell, an important factor in the electrolyte ionic conductivity which also dictates the resulting SEI composition. The utilization of electrochemically activated radicals is an interesting development towards in-situ studies with enhanced sensitivity.

Similarly, Overhauser DNP from metals is expected to be extremely beneficial. As currently only sodium and lithium metals were shown to lead to sensitivity gains, a broader study is needed to understand what degree of conductivity can result in DNP signal enhancements.

As for other applications in materials science, one can expect that with the expanding capabilities of DNP hardware, namely frequency agile microwave sources and pulse DNP methods, new applications will emerge. In the context of battery materials, many of the limitations associated with the electronic and magnetic properties of electrode materials may be less constricting. For example, pulse DNP can enable utilization of fast relaxing polarization sources and prevent some of the heating effects associated with conductive materials. Having a broader excitation range for the microwave would enable use of a wider range of paramagnetic metal ions for DNP.

Finally, one of the most exciting research directions in this area is the development of efficient DNP conditions at ambient temperature. This will open the way to in-situ approaches for studying battery materials in their native environments. This direction can be enabled by identifying DNP mechanisms functional at room temperature. In this regard, coherent polarization transfer schemes based on pulse DNP and relaxation mediated OE DNP seem to be very promising, giving rise to many intriguing fundamental and applied research questions.

## 7. Acknowledgments

Shira Haber was supported by a research grant from the Yotam Project and the Weizmann Institute Sustainability and Energy Research Initiative. We are thankful for support from the Israel Science Foundation (grant number 1580/17) and the European Research council (grant 803024 - MIDNP). This work was possible in part by the historic generosity of the Harold Perlman family.